\documentclass[preprint]{cimento}

\usepackage{amsmath}
\usepackage{amssymb}
\usepackage{epsfig}
\usepackage{graphicx}
\usepackage{multirow}
\usepackage{color}
\usepackage[normal]{subfigure}
\usepackage{rotating}

\newcommand{\bwt}{\begin{widetext}}
\newcommand{\ewt}{\end{widetext}}
\newcommand{\beq}{\begin{equation}}
\newcommand{\eeq}{\end{equation}}
\newcommand{\bdm}{\begin{displaymath}}
\newcommand{\edm}{\end{displaymath}}
\newcommand{\bea}{\begin{eqnarray}}
\newcommand{\eea}{\end{eqnarray}}

\def\eq#1{{Eq.~(\ref{#1})}}

\title{Recent developments in nonsupersymmetric $SO(10)$ unification}
\author{Luca Di Luzio\from{ins:x}\thanks{In collaboration with Stefano Bertolini (INFN \& SISSA, Trieste) and Michal Malinsk\'{y} (KTH, Stockholm). 
Poster session contribution given at IFAE2010, April 7-9 2010 Rome, Italy.}}
\instlist{\inst{ins:x} SISSA and INFN, Sezione di Trieste,
Via Bonomea 265, 34136 Trieste, Italy}
\PACSes{
}
\begin{document}

\maketitle

\begin{abstract}
I review the recent efforts in the search for a minimal and predictive nonsupersymmetric $SO(10)$ Grand Unified Thoery (GUT).
The outcome is the revival of a minimal scenario in which a long standing result, 
claiming the incompatibility between unification constraints and symmetry breaking dynamics, 
is now confuted by the implementation of the one-loop effective potential.
\end{abstract}

\section{The vacuum of the minimal nonsupersymmetric $SO(10)$ GUT}
\label{sec:intro}

In the last twenty years supersymmetric unification has played a major role, 
due to the success of weak scale supersymmetry (SUSY) in predicting gauge unification. 
On the other hand, SUSY might be broken at scales much higher than the TeV 
and it is not a mandatory ingredient for unification once we trade naturalness for predictivity. 
Without SUSY the dangerous $d=4$ and $5$ baryon number violating operators decouple from our 
low-energy world and the unification ansatz in nonsupersymmetric $SO(10)$ predicts the existence of intermediate scales, 
well separated from the GUT scale and in the ideal range for the 
generation of neutrino masses and for leptogenesis \cite{Chang:1984qr,Deshpande,Bajc:2005zf,Bertolini:2009qj}. 
For a recent analysis of fermion masses and mixings in nonsupersymmetric $SO(10)$, see ref.~\cite{Bajc:2005zf}.

However, devising a realistic and simple enough ordinary $SO(10)$ GUT is a rather non-trivial task  
and the main reason has to do with the structure of its minimal Higgs sector. 
The simplest conceivable (from a group-theoretically point of view) Higgs sector 
responsible for a full breaking of the GUT symmetry down to the Standard Model (SM) is given 
by a pair of Higgs multiplets: an adjoint, $45_H$, and a spinor, $16_H$.
A SM preserving breaking pattern is controlled then by two $45_H$ vacuum expectation values (VEVs) and one $16_H$ VEV.
Different configurations of the two adjoint VEVs preserve
different $SO(10)$ subalgebras,
namely,
$4_{C}\, 2_{L}\, 1_{R}$,
$ 3_{C}\, 2_{L}\, 2_{R}\, 1_{B-L}$,
$3_{C}\, 2_{L}\,1_{R}\,1_{B-L}$,
and the flipped or standard
$SU(5) \otimes U(1)$.
Except for the last case, the subsequent breaking to the SM is obtained
via the standard $SU(5)$ conserving $16_H$ VEV.

The phenomenologically favored scenarios allowed by nonsupersymmetric 
unification
correspond minimally to a two-step breaking along one of the following directions~\cite{Bertolini:2009qj}:
\begin{equation}
\label{chains}
SO(10)\stackrel{M_G}{\longrightarrow}
 3_{C}\, 2_{L}\, 2_{R}\, 1_{B-L}\stackrel{M_I}{\longrightarrow} \mbox{SM}
\,, \qquad
SO(10)\stackrel{M_G}{\longrightarrow}
4_{C}\, 2_{L}\, 1_{R}\stackrel{M_I}{\longrightarrow} \mbox{SM}
\,, 
\end{equation}
where the first breaking stage is driven by the $45_H$ VEVs, while the breaking to the SM at the 
intermediate scale $M_I$, several orders of magnitude below the unification scale $M_G$, is controlled
by the $16_H$ VEV.

Gauge unification, even without proton decay limits, excludes any
intermediate $SU(5)$-symmetric stages. On the other hand,
a series of studies in the early 1980's of the $45_H\oplus 16_H$ model 
\cite{Buccella:1980qb,Yasue,Anastaze:1983zk,Babu:1984mz} indicated that
the only intermediate stages allowed by the scalar sector dynamics were
the flipped $SU(5)\otimes U(1)$ for leading $45_H$ VEVs or the standard $SU(5)$ GUT
for  dominant $16_H$ VEV.
This observation excluded
the simplest $SO(10)$ Higgs sector from realistic consideration.

In ref.~\cite{Bertolini:2009es} we show that the exclusion of the breaking patterns
in \eq{chains} is an artifact of the tree-level potential. As a matter of fact,
some entries of the scalar hessian
are accidentally
over-constrained at the tree-level and a number of scalar interactions that, by a simple 
inspection of the relevant global symmetries and their explicit breaking, are expected to 
contribute to these critical states, are not effective at the tree-level.
On the other hand, once quantum corrections are taken into account,
contributions of $\mathcal{O}(M_G^2/16\pi^2)$ induced on these entries
open in a natural way
all group-theoretically allowed vacuum configurations.
Remarkably enough, the study of the one-loop effective potential
can be consistently carried out just for the critical tree-level hessian entries, 
while for all other states in the scalar spectrum, quantum corrections
remain perturbations of the tree-level results and do not affect
the discussion of the vacuum pattern.
These conclusions apply to any Higgs setting where the first step
of the $SO(10)$ gauge symmetry breaking is driven by the $45_H$ VEVs,
while the other Higgs representations control the
intermediate and weak scale stages. 

The results presented in ref.~\cite{Bertolini:2009es} open the path for reconsidering the minimal 
nonsupersymmetric $SO(10)$ GUT 
as a reference framework for unified model building.
Extending the Higgs sector to include one $10_{H}$ (together with either one ${16}_H$ 
or one ${126}_H$) provides the playground
for exploring the possibility of a realistic and predictive GUT,
along the lines of the recent efforts in the supersymmetric context.

\end{document}